\documentstyle[aaspp4]{article}

\lefthead{MENOU} 
\righthead{ADAFs AROUND WHITE DWARFs}

\def\gsim{\;\rlap{\lower 2.5pt
 \hbox{$\sim$}}\raise 1.5pt\hbox{$>$}\;}
\def\lsim{\;\rlap{\lower 2.5pt
   \hbox{$\sim$}}\raise 1.5pt\hbox{$<$}\;} 

\begin{document}

\title{Advection-Dominated Accretion Onto Weakly-Magnetized White
Dwarfs}

\author{Kristen Menou\altaffilmark{1}}

\affil{Princeton University, Department of Astrophysical Sciences,
Princeton NJ 08544, USA, kristen@astro.princeton.edu}

\altaffiltext{1}{Chandra Fellow}

\begin{abstract}

The boundary layers of weakly-magnetized white dwarfs (WDs) accreting
at rates $\lsim 10^{16}$~g~s$^{-1}$ are radially extended, hot,
optically-thin, and they advect some of their internally-dissipated
energy (Narayan \& Popham 1993). Motivated by this, I construct here
idealized spectral models of an Advection-Dominated Accretion Flow
(ADAF) around a WD, for application to quiescent Dwarf Novae (DN).
The Bremsstrahlung cooling of the gas in the ADAF, with temperatures
ranging from a few keV to a few tens of keV, can account for the X-ray
emission properties of quiescent DN.  If the energy advected by the
flow is thermalized in the WD atmosphere, the resulting emission from
the entire stellar surface (blackbody of temperature $T_{\rm eff} \sim
5$~eV) outshines the X-ray luminosity substantially.  This extreme-UV
component provides a flux in the 0.055-0.28~keV band which is
sufficient to power the strong HeII $\lambda4686$ emission lines of
quiescent DN by photoionization of the disk material. Reprocessing of
the ADAF X-ray emission by a cold outer thin disk could also lead to
an observable iron $K_{\alpha}$ fluorescence emission line, which can
be used to probe the geometry of the accretion flow. Existing
observational data indicate that the presence of ADAFs in quiescent DN
is not ubiquitous, while future observations, in particular with the
X-ray satellites {\it Chandra} and {\it XMM-Newton}, have the
potential to detect signatures of the hot flow in promising
candidates.

\end{abstract}


\section{Introduction}

Cataclysmic Variables (CVs) are binary stars in which a main-sequence
donor transfers mass via Roche-lobe overflow onto a White Dwarf
(WD). Many CVs are members of the subclass of transient systems called
Dwarf Novae (DN). DN regularly experience luminous outbursts during
which accretion onto the WD proceeds at a high rate. Most of the time,
however, DN are in quiescence, a phase during which the accretion rate
onto the WD (and the system luminosity) are much reduced (see Warner
1995 for a review).

DN in quiescence are well known sources of hard X-ray emission, with
typical luminosities $\sim 10^{30}-10^{32}$~erg~s$^{-1}$.  (C\'ordova
\& Mason 1983; Patterson \& Raymond 1985a).  Spectral fits to the
X-ray emission of these sources suggest a Bremsstrahlung origin, from
a gas with temperatures $\sim 2-20$~keV (Patterson \& Raymond 1985a;
Eracleous, Halpern \& Patterson 1991; Belloni et al. 1991; Yoshida,
Inoue \& Osaki 1992; Mukai \& Shiokawa 1993).

The X-ray emission of quiescent DN has commonly been attributed to
their boundary layer (BL), at the interface between the accreting WD
and the accretion disk surrounding it. Indeed, the BL can contribute
to a large fraction of the total luminosity of a system, typically
half of the accretion luminosity in the case of a non-rotating star
(see, e.g., Frank, King \& Raine 1992). At low accretion rates ($\lsim
10^{16}$~g~s$^{-1}$, such as in quiescent DN), the gas density in the
BL is low and the gas is unable to cool. This leads to a hot and
optically-thin BL which is a substantial source of hard X-ray emission
(Pringle \& Savonije 1979; Tylenda 1981; Patterson \& Raymond 1985a).

Detailed calculations by Narayan \& Popham (1993) show that the
optically-thin BLs of accreting WDs are also radially extended (on the
order of the WD radius in their models) and that they advect part of
the internally dissipated energy as a consequence of their inability
to cool. Partial radial pressure support by the hot gas also results
in sub-Keplerian rotation profiles in these solutions. These are all
properties that optically-thin BLs and Advection-Dominated Accretion
Flows (ADAFs; Ichimaru 1977; Rees et al. 1982; Narayan \& Yi 1994;
1995; Abramowicz et al. 1995) have in common, and it suggests that an
ADAF is a possible outcome of accretion at a low rate onto a
weakly-magnetized WD. In particular, if the radially extended BLs of
Narayan \& Popham (1993) were to extend, not to a stellar radius or
so, but to several tens of stellar radii, they would have properties
similar to those of ADAFs as they have been previously discussed in
the literature (see, e.g., Narayan, Mahadevan \& Quataert 1998 for a
review).

In this paper, I examine some consequences of the presence of an ADAF
around a WD. Advection can play an important role for the overall
energy budget and the emission spectrum of the accretion flow. In \S2,
I state the model assumptions, describe the ADAF models used for the
spectral predictions and explore their dependence on various model
parameters.  In \S3, models including a thin disk in the outer regions
of the flow are constructed for application to quiescent dwarf
novae. Finally, I discuss several implications and important
limitations of this work in \S4, before summarizing the main results
in \S5.

\section{ADAFs around White Dwarfs}

In the following, we assume for simplicity that the properties of an
ADAF around a WD can be described by simply truncating the ADAF
solutions that have developped for the case of accretion onto a BH. As
discussed in Appendix~A, it is not clear if this assumption is fully
valid.

The radius of a typical WD is $\sim 10^3 R_s$, where $R_s =3 \times
10^5 (M_{\rm WD}/M_\odot)$~cm is the Schwarzschild radius and $M_{\rm
WD}$ is the mass of the WD. At radii $\gsim 10^3 R_s$ in an ADAF,
electrons are non-relativistic (implying negligible synchrotron
emission) and the flow optical thickness to infinity is $\ll1$
(implying negligible Comptonization), so that Bremmstrahlung emission
is the only significant cooling mechanism for the flow. At these
radii, the ADAF is also essentially one-temperature because Coulomb
collisions transfer more efficiently the energy between the ions and
the electrons (and the non-relativistic electrons have the same
adiabatic index as the protons). Contrary to the case of accretion
onto a black hole, this makes the structure of the hot flow in this
context independent of the uncertain assumption of preferantial
viscous heating of the ions (Quataert \& Gruzinov 1999).  Note that,
in the following, we consider only accretion onto a weakly-magnetized
WD, in the sense that the stellar magnetic field is neglected in the
description of the accretion flow structure.

\subsection{Model Specifications}

The nature of the interface between the ADAF and the WD surface is
important for the emission models presented here. I make two further
simplifying assumptions regarding this interface: (1) the angular
velocity of the gas in the ADAF nearly matches that of the WD at the
stellar surface so that there is no additional shear (nor dissipation)
in this region, and (2) the gas carrying the energy advected by the
ADAF penetrates deep enough into the optically thick regions of the
stellar atmosphere that all this energy is thermalized before being
radiated away. The validity of these assumptions and the consequences
of relaxing them are discussed in more detail in Appendix~A.

With assumption (1), the steady-state energetics of the accretion flow
is equivalent to that described in Esin, McClintock \& Narayan (1997),
except for the effect of the presence of the stellar surface. The rate
of energy advection $L_{\rm adv}$ at the stellar surface and an
average advection parameter $f_{\rm adv}$ for the flow are obtained
from the difference between the local values of the viscous
dissipation rate and the radiative cooling rate, summed over the
entire radial extent of the hot flow. In cases where the ADAF
describes the corona of a thin accretion disk (i.e. with an accretion
rate varying with radius; see \S3), the energy required to evaporate
the gas from the disk to the corona is also taken into account in the
energy budget as a sink for the ADAF (see Esin et al. 1997 and \S3 for
details).

The ADAF models used here for the spectral predictions are similar to
the models described in Narayan et al. (1998; see also Quataert \&
Narayan 1999b), except for the emission from the stellar surface.  The
reemission of the energy advected by the flow is modeled as a
single-temperature blackbody of temperature:
\begin{equation}
T_{\rm eff}=\left( \frac{L_{\rm adv}}{4 \pi f \sigma R_{\rm WD}^2}
\right)^{1/4},
\label{eq:one}
\end{equation}
where $f$ is the fraction of the stellar surface that is emitting
($f=1$ is assumed in what follows). The interaction of this extra
component of the radiation field with the gas in the ADAF is
calculated self-consistently (see Menou \& McClintock 2000 for
details), but is found to be unimportant in all the models presented.

Unless otherwise specified, the values adopted for the ADAF parameters
are $\alpha_{\rm ADAF}=0.2$ (viscosity parameter), $\beta=6$ (ratio of
gas to magnetic pressure), $\delta=10^{-2}$ (fraction of direct
electron viscous heating; again, the precise value of this parameter
is unimportant for the present models), $\gamma=1.636$ (adiabatic
index of the fluid, including contributions from the particles,
magnetic field and turbulence; see Quataert \& Narayan 1999a), $p=0$
(no wind) and $i=60^o$ (inclination). For the accreting WD, a mass
$M_{\rm WD}=1.2~M_\odot$ and a radius $R_{\rm WD}=5 \times 10^8 cm$
are generally assumed. Note that accretion rates are often expressed
in Eddington units ($\dot m=\dot M/\dot M_{\rm Edd}$, with $\dot
M_{\rm Edd}=1.39 \times 10^{18}~(M_{\rm WD}/M_\odot)$~g~s$^{-1}$).
All the models described in this section have ADAFs extending from the
surface of the WD up to $10^4~R_s$. The luminosities predicted by the
models are calculated by numerically integrating the spectral energy
distributions in the appropriate energy range. X-ray luminosities are
given for the $0.1-3.5$~keV band; this corresponds to the energy range
covered by the X-ray satellite {\it Einstein}, which has collected the
most extensive X-ray data set on DN to date (see, e.g., Eracleous et
al. 1991).

\subsection{Emission spectrum}
 
\begin{figure}
\plotone{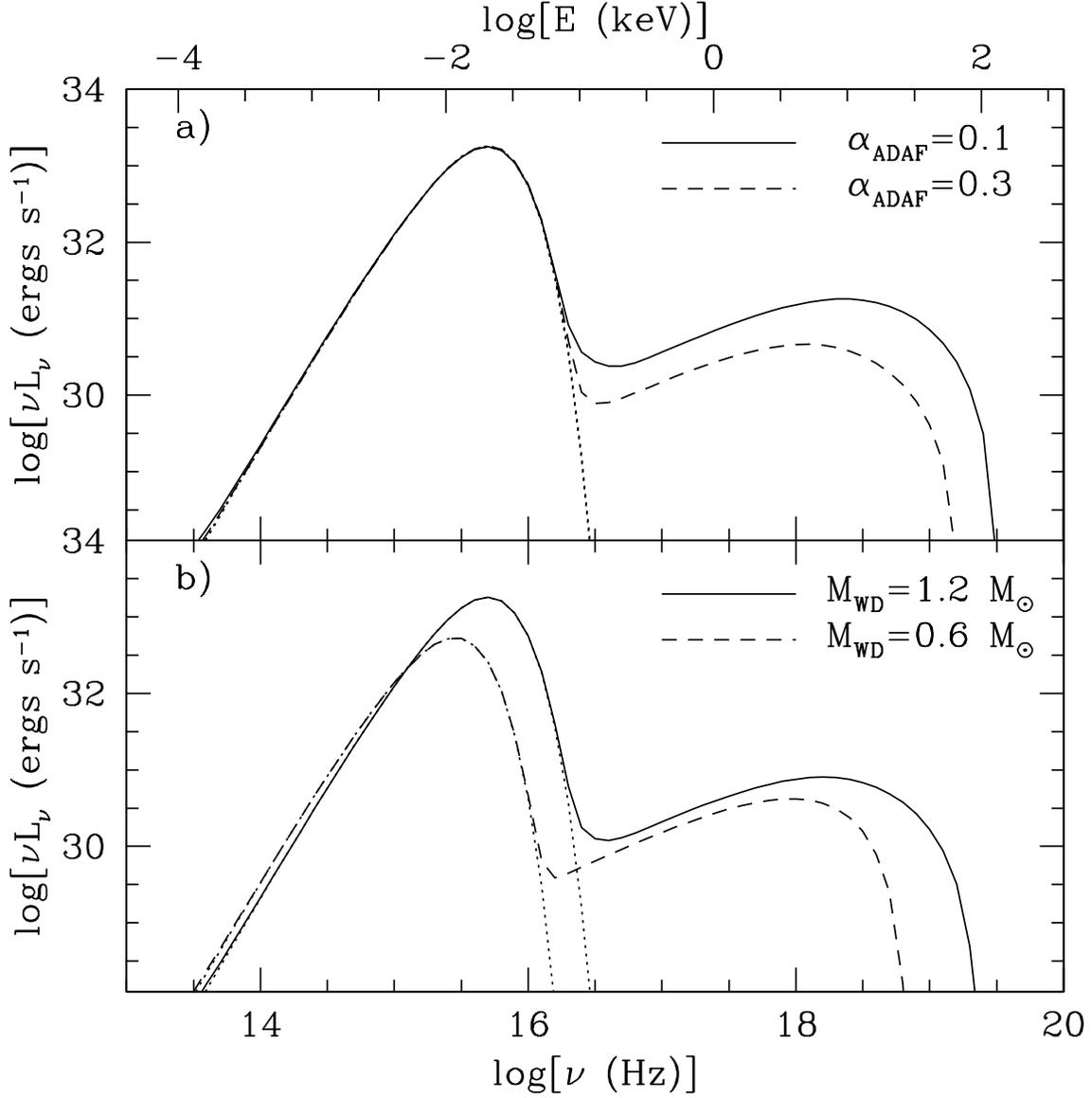}
\caption{The emission spectrum of an ADAF around a WD. The dotted line
corresponds to the reemission of the energy advected by the flow from
the photosphere of the accreting WD (assuming thermalization; see
text). X-rays are due to Bremmstrahlung cooling of the hot gas in the
flow. (a) Influence of the viscosity parameter $\alpha_{\rm ADAF}$ on
the model predictions ($M_{\rm WD}=1.2 M_\odot$). (b) Influence of the
mass -- and compactness -- of the accreting WD on the model
predictions ($\alpha_{\rm ADAF}=0.2$; see text for details).
\label{fig:one}}
\end{figure}

Figure~\ref{fig:one} and~\ref{fig:two} show the emission spectra of
ADAFs around WDs for various values of the model parameters. All these
models are characterized by a two-component spectrum. The first
component, energetically dominant in the hard X-rays, is the
Bremmstrahlung emission of the hot gas in the ADAF. The second
component, energetically dominant in the EUV, is the accretion-powered
emission from the WD surface. This EUV component dominates the ADAF
X-ray emission because of the importance of energy advection: most of
the energy dissipated in the flow is radiated at the WD surface.  For
simplicity, we assume that the X-ray photons emitted by the flow that
reach the WD surface are absorbed and reemitted in the EUV
component. The energy deposited by these photons at the WD atmosphere
is generally negligible compared to the energy advected by the flow.
The EUV radiation from the WD surface is found to escape to infinity
without significantly interacting with the gas in the ADAF.

Note that the energetics of the EUV and X-ray components in the
various models described below can be simply understood as follows. In
first approximation, the luminosity of the EUV component from the WD
is $L_{\rm EUV} \propto \dot m$. The X-ray luminosity of the ADAF, on
the other hand, is due to Bremstrahlung emission, which scales as the
gas density squared, so that $L_{\rm X} \propto \dot m^2/\alpha_{\rm
ADAF}^2$ (Narayan et al. 1998). The spectral models are characterized,
to first order, by the ratio
\begin{equation}
\frac{L_{\rm X}}{L_{\rm EUV}} \propto \dot m/\alpha_{\rm ADAF}^2.
\end{equation}

\subsection{Influence of $\alpha_{\rm ADAF}$}

Figure~\ref{fig:one}a shows the emission spectra of two ADAF models
with a same accretion rate $\dot m = 7 \times 10^{-3}$ but values of
the viscosity parameter $\alpha_{\rm ADAF} = 0.1$~and
$0.3$.\footnote{In the model with $\alpha_{\rm ADAF}=0.1$, the value
of the parameter $\beta$ is 10, while $\beta=3$ in the model with
$\alpha_{\rm ADAF}=0.3$. This is to represent the larger viscosity
expected when the magnetic pressure in the flow is larger, as
suggested by MHD simulations of accretion disks (see Narayan et
al. 1998 for details).}  In the model with $\alpha_{\rm ADAF}=0.1$,
the gas in the ADAF has a smaller radial velocity and a larger density
(Narayan \& Yi 1994), responsible for the more important X-ray
Bremmstrahlung emission than in the model with $\alpha_{\rm
ADAF}=0.3$. The EUV luminosity is very similar in both models because
most of the dissipated energy is advected and radiated by the WD
photosphere in both cases.  In the model with $\alpha_{\rm ADAF} =
0.1$, the advection parameter is $f_{\rm adv}=0.437$ and the X-ray
luminosity (0.1-3.5~keV) is $\simeq 2 \times 10^{31}$~erg~s$^{-1}$.
In the model with $\alpha_{\rm ADAF} = 0.3$, the advection parameter
is $f_{\rm adv}=0.937$ and the X-ray luminosity is $\simeq 7.5 \times
10^{30}$~erg~s$^{-1}$. In both models, the total EUV luminosity is
$\simeq 2.5 \times 10^{33}$~erg~s$^{-1}$, at a temperature $T_{\rm
eff} \simeq 5.3$~eV.

\subsection{Influence of $M_{\rm WD}$}

Figure~\ref{fig:one}b shows the emission spectra of two ADAF models,
one with a WD of mass $M_{\rm WD}= 1.2 M_\odot$ and radius $R_{\rm
WD}=5 \times 10^8$~cm as usual, and with $\dot m = 7 \times 10^{-3}$;
in the second model, the WD has a mass $M_{\rm WD}= 0.6 M_\odot$, a
radius $R_{\rm WD} = 8.5 \times 10^{8}$ cm, and the accretion rate has
been doubled to $\dot m = 1.4 \times 10^{-2}$ to allow for a
comparison at the same physical accretion rate $\dot M = \dot m \dot
M_{\rm Edd} \simeq 10^{16}$~g~s$^{-1}$. The smaller mass and
compactness of the WD in the model with $M_{\rm WD}=0.6 M_\odot$ are
responsible for the reduced and cooler photospheric WD emission, and
the softer X-ray emission (because of a smaller range of temperatures
in the ADAF; see Fig.~\ref{fig:two}b). In the model with $M_{\rm WD}=
1.2 M_\odot$, the advection parameter is $f_{\rm adv}=0.85$, the X-ray
luminosity (0.1-3.5~keV) is $\simeq 1.1 \times 10^{31}$~erg~s$^{-1}$
and the total EUV luminosity is $\simeq 2.5 \times
10^{33}$~erg~s$^{-1}$, at a temperature $T_{\rm eff} \simeq
5.3$~eV. In the model with $M_{\rm WD}= 0.6 M_\odot$, the advection
parameter is $f_{\rm adv}=0.665$, the X-ray luminosity is $\simeq 7.3
\times 10^{30}$~erg~s$^{-1}$ and the total EUV luminosity is $\simeq
7.3 \times 10^{32}$~erg~s$^{-1}$, at a temperature $T_{\rm eff} \simeq
3$~eV.

\subsection{Influence of $\dot m$}

\begin{figure}
\plotone{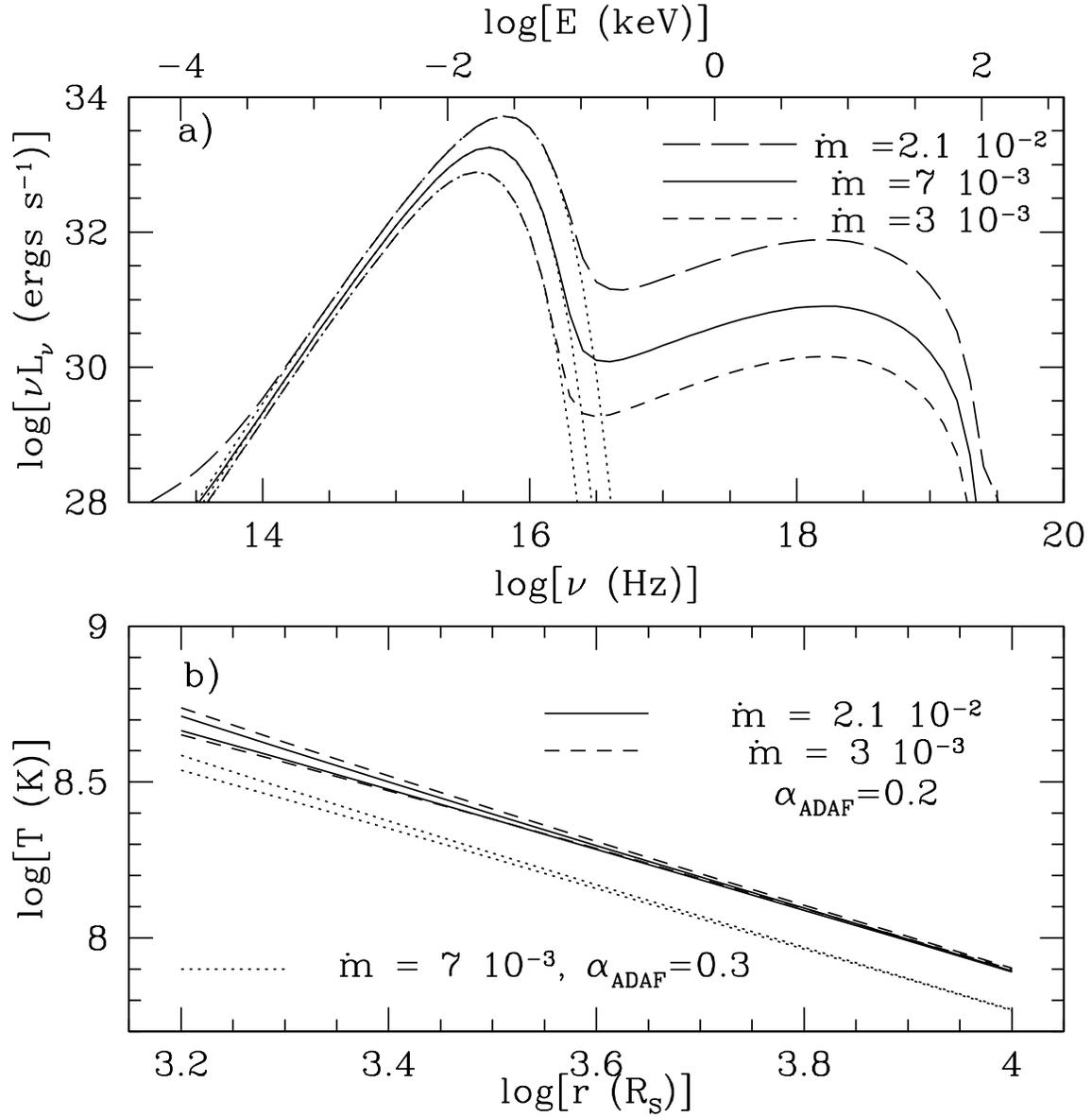}
\caption{(a) Same as Fig.~1, but shows the influence of the accretion
rate $\dot m$ on the model predictions. (b) Radial temperature
profiles for the ions (upper) and the electrons (lower) in three
models with various parameters.
\label{fig:two}}
\end{figure}

Figure~\ref{fig:two}a shows three models with accretion rates $\dot m
= 2.1 \times 10^{-2}$, $\dot m = 7 \times 10^{-3}$ and $\dot m = 3
\times 10^{-3}$.  At higher accretion rates, the density in the ADAF
is larger and the cooling is more efficient, so that the ratio of EUV
to X-ray luminosity is reduced. The advection parameter in these
models is $f_{\rm adv}=0.63$, $0.85$ and $0.93$, respectively.  The
corresponding X-ray luminosity ($0.1-3.5$~keV) for each model is
$\simeq 10^{32}$, $1.1 \times 10^{31}$ and $2 \times
10^{30}$~erg~s$^{-1}$, respectively. The corresponding total EUV
luminosity for each model is $\simeq 7.2 \times 10^{33}$~erg~s$^{-1}$
(effective temperature $T_{\rm eff} \simeq 6.9$~eV), $2.5 \times
10^{33}$~erg~s$^{-1}$ ($T_{\rm eff} \simeq 5.3$~eV) and
$10^{33}$~erg~s$^{-1}$ ($T_{\rm eff} \simeq 4.3$~eV), respectively.

Figure~\ref{fig:two}b shows that the ion and electron temperatures in
the accretion flow are almost independent of the accretion rate $\dot
m$ (contrary to the two-temperature case; see Esin et al. 1997). The
difference in the temperature profile for the model with $\alpha_{\rm
ADAF}=0.3$ can be attributed to the smaller value of $\beta$, which
modifies the relative importance of the gas and magnetic pressure in
the flow.  The temperature profile does not depend on $M_{\rm WD}$, in
the sense that if one were to superpose the temperature profiles of
the two models shown in Fig.~\ref{fig:one}b, they would be nearly
indistiguishable.  The shorter radial extent of the ADAF in the model
with $M_{\rm WD}= 0.6 M_\odot$, however, implies a cooler accretion
flow, with a maximum temperature $\log [T(K)] \simeq 8.2$ at the WD
surface. The range of gas temperatures in the ADAF (from a few keV to
a few tens of keV) depends therefore primarily on the radial extent of
the hot flow.

\section{Models with an Outer Thin Disk}

More realistic models for quiescent DN must include the presence of a
thin disk in the outer regions of the accretion flow. According to the
disk instability model (DIM), this disk is responsible for the
outbursts of DN, by storing mass during quiescence and accreting it
suddenly during outburst under the action of a thermal-viscous
instability (see Cannizzo 1993 for a review of the DIM).

Ignoring the exact structure of the transition region from the disk to
the ADAF, I adopt here for simplicity the prescription of Esin et
al. (1997). Beyond a transition radius ($R_{\rm trans}$) to be
defined, the mass accretion rate in the ADAF (= the disk corona)
declines as $R_{\rm trans}/R$ . This corresponds, as one goes in, to a
gradual evaporation of the disk material into the corona, up to the
point where, inside $R_{\rm trans}/R$, accretion occurs exclusively
via the ADAF (Esin et al. 1997).

The disk reflection component and its iron $K_{\alpha}$ fluorescence
line are also modeled following Esin et al. (1997). Angle-averaged
energy-dependent Green's functions are used to predict the intensity
of the line which depends on the supply of hard X-ray photons from the
ADAF irradiating the disk (Lightman \& White 1988; George \& Fabian
1991). The disk is supposed to contain neutral material with a cosmic
abundance of $3.3 \times 10^{-5}$ iron atoms per hydrogen atom
(Morrison \& McCammon 1983). The line profile is not modeled in detail
here. However, the contribution to the line emission of each disk
annulus, at a radius $R$ from the WD, is given the proper fractional
Doppler width $\sin i/(2R/R_s)^{1/2}$ ($i$ is the disk inclination to
the line of sight) and is added to the continuum emission in the
proper energy bin of the computed spectrum.  Note that the simple
prescription used here for the variation of the accretion rate with
radius beyond $R_{\rm trans}$ does not necessarily guarantee that the
disk material is neutral everywhere, as expected in the DIM (Wheeler
1996). It is not important for our purpose, however, because we are
not interested in the disk emission properties, but instead explore
the effect of varying its radial extent on the intensity of the
reflection component, assuming that the disk is neutral as the DIM
would predict.

\subsection{The reflection component}

\begin{figure}
\plottwo{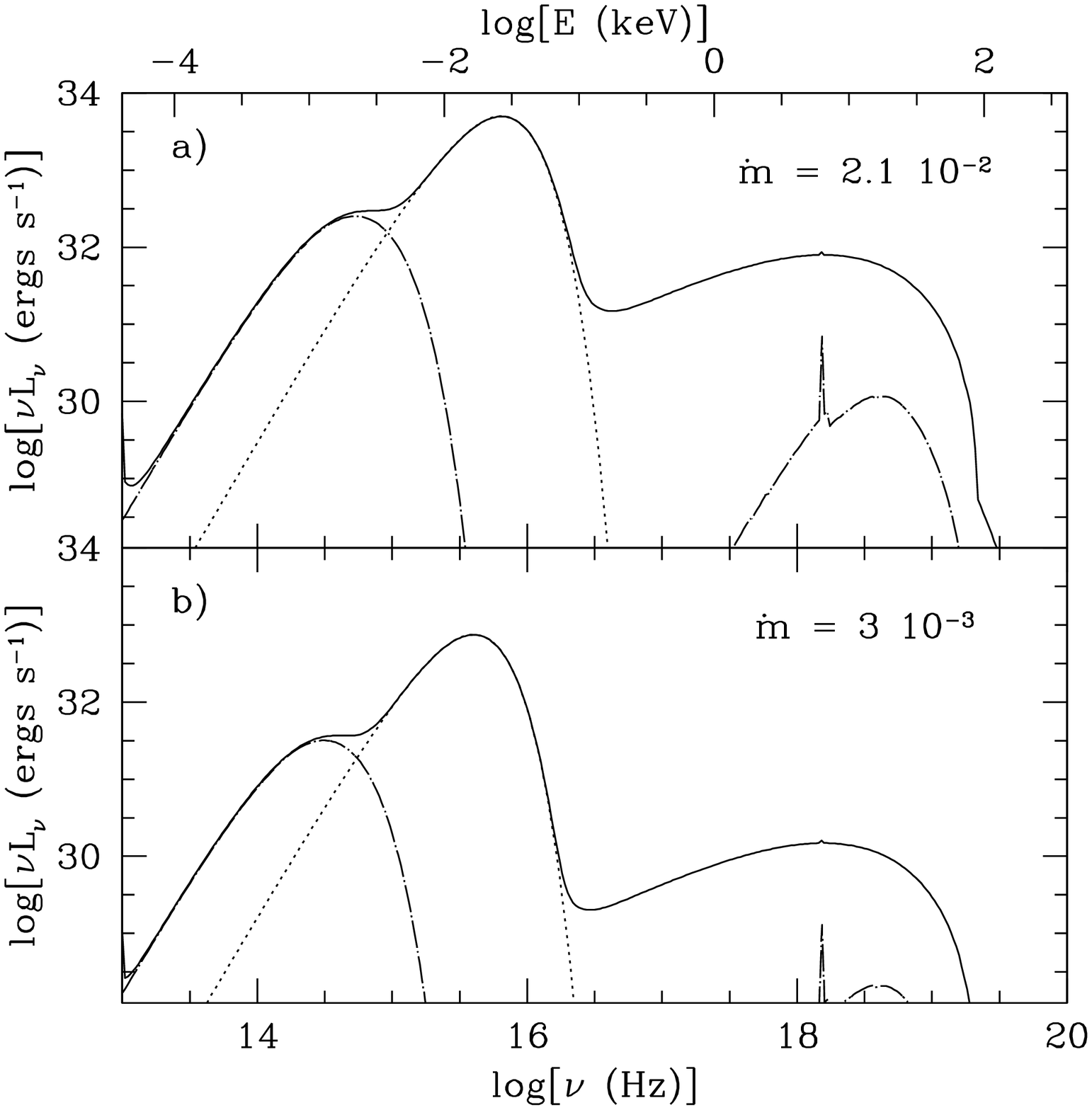}{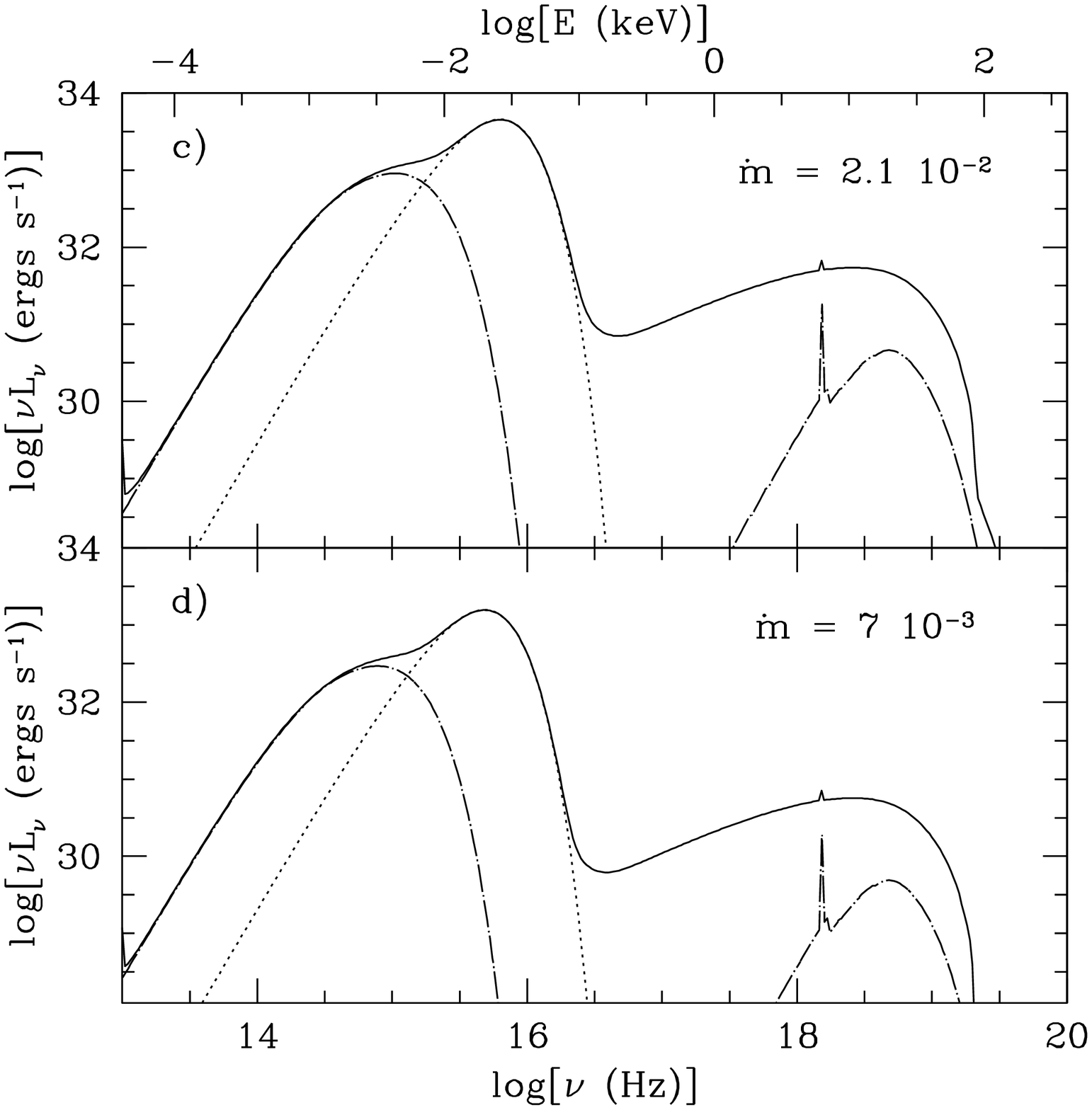}
\caption{Spectral models for quiescent Dwarf Novae, including the
contribution from an outer thin disk and the X-ray emission that it
reprocesses (long dashed-dotted), the EUV emission resulting from the
advection of energy onto the central WD (dotted) and the ADAF
Bremmstrahlung emission. Models (a) and (b) have a disk extending from
$10^{4} R_s$ to $10^{5} R_s$, while models (c) and (d) have a disk
extending further in, from $10^{3.5} R_s$ to $10^{5} R_s$. The iron
$K_\alpha$ emission line at 6.4~keV is a probe of the geometry of the
accretion flow.
\label{fig:three}}
\end{figure}

Figure~\ref{fig:three}a shows the spectra predicted for a thin disk +
ADAF model, under various assumptions for the accretion rate $\dot m$
and the disk radial extent. An inclination $i=60^o$ is assumed in all
four models.  The solid line represents the total emission from the
accretion flow, including the EUV contribution from the WD surface
(dotted) and the contribution from the disk (long-dashed-dotted). The
disk reflection component peaks at X-ray frequencies, while the disk
intrinsic emission peaks at optical frequencies (again, this component
is not supposed to represent the actual emission of a quiescent disk
according to the DIM -- see above). In all cases, the disk reflection
component contributes only to a small fraction of the continuum X-ray
flux, as expected. Depending on the geometry of the accretion flow,
however, an iron $K_\alpha$ emission line is present and is
potentially observable on top of the X-ray continuum.

Figure~\ref{fig:three}a~and~b show two models with a value of the
transition radius $R_{\rm trans}=10^4 R_s$ and a disk (+ ADAF-corona)
extending up to $10^5 R_s$ (the precise value of this maximum radius
is unimportant for the spectral predictions discussed
here). Figure~\ref{fig:three}c~and~d show two models with a disk
extending further in, down to $R_{\rm trans}=10^{3.5} R_s$. In each
case, models for two values of the accretion rate $\dot m$ are shown.

The luminosity in the iron $K_\alpha$ fluorescence line at 6.4~keV
measures the amount of photons absorbed by the disk at energies $>
7.1$~keV, corresponding to the iron K-shell absorption edge seen in
each of the reflection components in
Fig.~\ref{fig:three}. Consequently, the line intensity mainly depends
on the accretion rate in the ADAF (for the overall X-ray luminosity)
and the location of the disk relative to the hottest regions of the
flow (to guarantee the absorption by the disk of a large amount of
hard enough photons).  Fig~\ref{fig:two}b shows that temperatures
above $\sim 10^8$~K in the ADAF are found only inside $R \simeq 10^4
R_s$, so that little or no $K_\alpha$ line emission is expected for
disks extending beyond this limit. This is confirmed by
Fig.~\ref{fig:three}a~and~b which show rather weak line emission in
the reflection component, and no significant emission on top of the
continuum, for two different values of $\dot m$. On the other hand,
Fig.~\ref{fig:three}c~and~d (corresponding to $R_{\rm
trans}=10^{3.5}R_s$) show stronger iron fluorescence lines, which can
be seen on top of the Bremsstrahlung continuum emission from the hot
flow.  Obviously, the lines would be even stronger if the disk was
extending further in. The strength of this line is therefore a probe
of the geometry of the accretion flow in the system.  For $i=60^o$ and
$R=10^{3.5}R_s$, resolving the width of the iron lines shown in
Fig.~\ref{fig:three}c~and~d requires a spectral resolution $\simeq
92$, so that the resolving power of order $200$ that can be reached
with {\it Chandra} around 7 ~keV (with HETG) could allow detailed
diagnostics of the line-emitting material if such a line is present
and strong enough.

Note that ADAFs around low-mass WDs (say $<0.5 M_\odot$ or so) are not
expected to be surrounded by a gas with temperatures much in excess of
$10^8$~K because of the large size of the compact object
(Fig.~\ref{fig:two}b). DN containing low-mass WDs are therefore not
expected to show significant iron fluorescence line emission no matter
how extended their disks are.

\subsection{Explaining the HeII emission lines}

A likely origin for the strong HeII $\lambda4686$ emission lines shown
by many DN is photoionization of the disk material by very soft X-rays
in the energy range 55-280~eV (Patterson \& Raymond 1985b).  However,
there is no direct detection of these irradiating photons by either
{\it ROSAT} (soft X-rays) or {\it IUE} (UV).  This led Vrtilek et
al. (1994) to propose that these photons could originate from an EUV
component, such as blackbody emission with a temperature $\sim
10-20$~eV. This makes the advection-powered EUV emission from a WD
surrounded by an ADAF a good candidate for explaining the strength of
the HeII emission lines in quiescent DN.

A detailed calculation of the strength of these lines for a given
accretion model requires the knowledge of the amount of photons with
appropriate energies which are absorbed by the disk. The result of
this calculation depends on the the disk geometry (which is not
described properly in the present models given the arbitrary profile
assumed for $\dot m$ in the disk) and, more importantly in the present
case, on the assumption of exact reemission as a single-temperature
blackbody for the advected energy, which has a strongly peaked
contribution at low energies in the 0.055-028~keV range (see, e.g.,
Fig.~\ref{fig:one}).  This means that even a slight deviation from the
strict blackbody emission hypothesis could largely influence the
predictions for the line strength. Consequently, I use here, instead
of a detailed calculation, the simple argument of Patterson \& Raymond
(1985b; see also Vrtilek et al. 1994) to show that the spectral models
of Fig.~\ref{fig:three} can account for the observed luminosities of
the HeII lines in quiescent DN.

Following Patterson \& Raymond (1985b), the luminosity of a system in
the HeII $\lambda4686$ lines is estimated as
\begin{equation}
L_{\lambda4686}=(0.26)(\eta) \left( \frac{2.65~{\rm eV}}{130~{\rm eV}}
\right) \times L_{0.055-0.28}= 5.3 \times 10^{-4}~ L_{0.055-0.28},
\label{eq:irrad}
\end{equation}
where 0.26 is the fraction of He recombinations resulting in photons
at $\lambda = 4686 \AA$, $\eta=0.1$ is a geometrical factor
corresponding to the fraction of the irradiating source luminosity
that is actually absorbed by the disk,\footnote{The value $\eta=0.1$
used by Patterson \& Raymond (1985b) is based on a geometric model of
the central X-ray source. The authors quoted an uncertainty of a
factor 3 in either direction for $\eta$. This same value should be
appropriate for the order-of-magnitude estimates made here.} $130$~eV
is an assumed energy for a typical ionizing photon and
$L_{0.055-0.28}$ is the luminosity of the irradiating source (the WD
here) in the $0.055-0.28$~keV energy range.  Eq.~(\ref{eq:irrad}) has
been used by Patterson \& Raymond (1985b; see also Vrtilek et
al. 1994) to argue for the existence of a yet unseen soft source of
ionizing photons.  Note that because the EUV emission from the WD
surface in the energy range 0.055-0.28~keV strongly peaks at low
energies in the present models, the typical energy of ionizing photons
could reasonably be estimated as being less than $130$~eV. This, and
possible deviations from a pure blackbody emission, could result in
somewhat larger values for $L_{\lambda4686}$ than estimated below.

For the three models shown in Fig.~\ref{fig:two}a (or, equivalently,
the models with the same $\dot m$ in Fig.~\ref{fig:three}), the value
of $L_{0.055-0.28}$ are $\simeq 1.4 \times 10^{33}$~erg~s$^{-1}$
($\dot m=2.1 \times 10^{-2}$), $1.1 \times 10^{32}$~erg~s$^{-1}$
($\dot m=7 \times 10^{-3}$) and $10^{31}$~erg~s$^{-1}$ ($\dot m= 3
\times 10^{-3}$). This corresponds, according to Eq.~(\ref{eq:irrad}),
to luminosities $L_{\lambda4686} \simeq 7.4 \times
10^{29}$~erg~s$^{-1}$, $5.8 \times 10^{28}$~erg~s$^{-1}$ and $5.3
\times 10^{27}$~erg~s$^{-1}$, respectively. This is within the range
of observed values for quiescent DN with accretion rates $\lsim
10^{16}$~g~s$^{-1}$ (Patterson \& Raymond 1985b), corresponding to
$\dot m \lsim 10^{-2}$.  The models presented here satisfy the
criterion for the postulated missing soft component (Patterson \&
Raymond 1985b): an emission which provides a luminosity equal or
superior to that in the 0.1-3.5~keV range and which is suitable for
concentrating a large fraction of the luminosity in the 55-280~eV
energy range. Note that for small WD masses, the contribution of the
EUV component in the 55-280~eV energy band can be much reduced because
$R_{\rm WD}$ is larger and the EUV emission is then softer. This,
again, depends critically on the assumption of exact reemission like a
single-temperature blackbody.

The success of the proposed accretion models in providing an
explanation to both the X-ray emission and the strong HeII
$\lambda4686$ emission lines of quiescent DN must be moderated,
however, by noting that existing observations indicate that these
models cannot be applied to the entire population of quiescent DN (see
below).

\section{Discussion}

I have constructed idealized models of Advection-Dominated Accretion
Flows (ADAFs) around WDs, for application to quiescent DN. In the more
elaborate models, an ADAF is present in the inner regions of the flow,
while accretion proceeds via a thin disk (plus a corona, treated as an
ADAF here) in the outermost regions. According to the Disk Instability
Model (DIM), the thin disk is a reservoir a mass used to power the
dwarf nova outbursts (see, e.g., Cannizzo 1993). These models are
inspired from the models of Narayan, Barret \& McClintock (1997; see
also Lasota, Narayan \& Yi 1996) for quiescent Black Hole (BH) Soft
X-ray Transients. If the presence of such a two-component accretion
flow is established in some quiescent DN, it may be possible to
interpret the various spectral states of these systems as changes in
the geometry of the two-component accretion flow, by analogy with the
models for BH transients (see Esin et al. 1997; 1998).

However, the models for quiescent DN differ significantly from those
of BH transients for at least two reasons. First, the properties of
the ADAF are different here because the radial extent of the hot flow
is reduced around a WD (an object much less compact than a
stellar-mass BH). As a consequence, the gas in the hot flow is
one-temperature (because of efficient Coulomb coupling at lower
temperatures) and the property of energy advection is independent of
the assumption of preferential heating of the ions by the viscous
dissipation (and poor energy exchange between the ions and the
electrons) made for two-temperature ADAFs.

Second, and more importantly, while the energy advected by the flow is
lost through the event horizon in the BH case, it must be reradiated
from the hard surface of the WD in the DN case. Assuming that this
energy is thermalized in the WD atmosphere, I have shown that the
resulting EUV emission from the WD surface could explain the puzzling
strength of HeII $\lambda 4686$ emission lines of quiescent DN as due
to disk irradiation, while the X-ray properties of the same systems
could be explained by the X-ray emission from the ADAF itself.  The
strong EUV component is not unique to ADAF models but could also be a
consequence of substantial energy advection in the BL solutions of
Narayan \& Popham (1993) if, again, this energy is thermalized before
being radiated away. A direct test of the existence of this strong EUV
component, as a possible signature of energy advection in the flow
surrounding the WD, would be quite valuable.

It may not be easy to discriminate, based on observations, between the
BL solutions of Narayan \& Popham (1993) and a radially extended ADAF
around a WD. They are both sources of hard X-ray emission and energy
advection. The BL solutions are characterized by a small radial extent
compared to the ADAF models presented here and higher gas densities:
$\rho \sim 10^{-10.5}$~g~cm$^{-3}$ in the model~d of
Fig.~\ref{fig:three}, while $\rho \sim 10^{-9}$~g~cm$^{-3}$ in a
comparable model of Narayan \& Popham (1993). The emission spectrum of
the BL solutions has not been calculated by Narayan \& Popham (1993),
but it is clear that the higher densities counterbalance the smaller
radial extent for the total X-ray emission. If the ADAFs were much
less radially extended than assumed in the proposed models (say a few
stellar radii at most), they could not produce X-ray emission at the
level observed in quiescent DN because of their low densities.

The proposed models for quiescent DN (with a inner hot flow replacing
the traditional thin disk) share some properties with the coronal
``siphon flow'' model of Meyer \& Meyer-Hofmeister (1994). An
important difference, however, is the presence in their model of a
wind from the hot flow, which is absent in the ADAF models presented
here. It has been proposed that ADAFs could have winds (Narayan \& Yi
1994; Blandford \& Begelman 1999) able to carry away some of the
dissipated energy that is not radiated by the flow. This could have
consequences for the emission spectrum of a system (in particular a
reduction of the EUV component originating at the WD surface) that
were not considered here.  Interestingly, Meyer \& Meyer-Hofmeister
(1994) proposed that heat conduction between the hot flow and the cold
gas in the disk is responsible for the continuing ``evaporation'' of
the cold disk material into the hot coronal flow. This process could
operate equally well in the present context (ADAF + thin disk) and may
be responsible for a transition from the BL solutions with a modest
radial extent of Narayan \& Popham (1993) to a more radially extended
ADAF as considered in the present study.

The presence of an inner, extended hot flow in quiescent DN could also
influence the outburst properties of these systems, as first
emphasized by Meyer \& Meyer-Hofmeister (1994; see also Liu, Meyer \&
Meyer-Hofmeister 1997). These authors pointed out that the presence of
this low density region could be responsible for the well-known UV
delay of DN (see, e.g., Warner 1995; Mauche 1996), by ensuring that,
during the rise to outburst, the disk has to fill in the inner regions
of the flow before a substantial amount of mass is accreted onto the
WD (Meyer \& Meyer-Hofmeister 1994; Liu et al.  1997; Hameury, Lasota
\& Dubus 1999). An analoguous situation has been suggested for the BH
X-ray transient GRO J1655-40 because of a delay of the rise of the
X-ray flux relative to the optical flux during the April 1996 outburst
of this system (Orosz et al. 1997); this delay has been modeled
successfully by Hameury et al. (1997) with a two component, inner ADAF
+ outer thin disk, time-dependent accretion model.

Observations show that some quiescent DN are more likely to harbor an
ADAF than others.  Of the five DN observed in quiescence by the
Hopkins Ultraviolet Telescope (Long 1996; spectral range $\sim
830-1860 \AA $), two have shown the expected spectral UV signatures
from a WD atmosphere (in particular a broad Ly$\alpha$ absorption
feature and narrow metal absorption lines): U Gem and VW Hyi.
However, the remainning three (WX Hyi, Yz Cnc and SS Cyg) show strong,
blue continua with no evidence for the presence of a normal WD
atmosphere (broad emission lines and no evidence for the expected
Ly$\alpha$ absorption feature).  These three sources with strong blue
excesses are ideal candidates for the presence of an ADAF because a
blue UV continuum is expected if the WD atmosphere is heated up to a
several $10,000$~K by a substantial amount of energy advection in the
surrounding hot flow.  That two out of the five observed quiescent DN
possess the signature expected from a ``naked'' WD shows the large
diversity of these systems and suggest that ADAF models do not apply
to the entire DN population. On physical grounds, a spread in the
observational properties of quiescent DN could be due, for instance,
to differences in the WD rotation rate, the surface magnetic field
strength or the radial extent of the hot flow for the various systems.
Note that observations during quiescence suggest the existence of an
``accretion belt'' on the central WD in the DN VW HYi (and perhaps U
Gem; see Sion 1999 for a review). This accretion belt, hotter than the
rest of the WD surface, is supposedly the result of accretion at a
high rate via a disk during DN outbursts (Kippenhahn \& Thomas 1978).
The presence of such accretion belts and of ADAFs in quiescence are
not mutually exclusive.

Similarily, existing X-ray observations put strong constraints on the
presence of ADAFs in some quiescent DN.  X-ray eclipses during the
quiescence of the dwarf novae HT Cas (Wood et al. 1995; Mukai et
al. 1997) and OY Car (Pratt et al. 1999) show that the size of the
X-ray emitting region is at most comparable to the WD radius, and
Szkody et al. (1996) claim that the X-ray emitting region in U Gem
during quiescence is small as well. These observations appear
inconsistent with the emission from a radially extended ADAF because
the bulk of the Bremmstrahlung emission, with an emissivity $\propto
\rho^2 T^{-1/2}$ and for the profiles $\rho \propto R^{-3/2}$ and $T
\propto R^{-1}$ in an ADAF, originates from the outermost regions of
the flow (see also Quataert \& Narayan 1999b). This indicates, again,
that ADAF models do not apply to the entire DN population, or perhaps,
that if a hot flow is present in these systems, it has a structure
different from an ADAF (for instance with a steeper density profile,
which is contrary to what is expected for an ADAF with wind; Blandford
\& Begelman 1999). van Teeseling, Beurmann \& Verbunt (1996) claim
that the X-ray emitting region in non-magnetic CVs is close to the WD
from an anticorrelation of the soft X-ray luminosity and the
inclination of several of these sources. This cannot be used to
strongly argue against the extended ADAF hypothesis because even a
slight deviation from the assumed blackbody-type emission of the
advected energy could make this component, originating close to the
WD, dominate the soft X-ray flux of the system.

There are other possible consequences of the presence of an ADAF in
quiescent DN which have not been considered in any detail here. The
presence of such a hot flow during the long phases of quiescence could
influence the time-averaged rotation rate of the weakly-magnetized
WD. The star accretes from a Keplerian disk during outburst, but could
also accrete from a substantially sub-Keplerian flow during
quiescence. At late evolutionary stages, the star could therefore spin
at an equilibrium rate which is in between the breakup value
($\Omega_K(R_{\rm WD})$) and the rotation rate of the gas in the ADAF
($\sim 0.2-0.3 \Omega_K$), rather than close to the breakup rate if it
accretes all the time from a disk.

The presence of an ADAF in a quiescent DN can be directly tested by
the detailed spectroscopic X-ray observations which have become
possible with the satellites {\it Chandra} and {\it XMM}. The
continuum emission of the ADAF is characteristic of a gas with
temperatures in the range of a few keV to a few tens of keV, depending
on the radial extent of the hot flow.  The hot gas in the ADAF is also
ideally suited for emitting thermal X-ray lines (Narayan \& Raymond
1999). The detectability of these lines and their role as possible
diagnostics of the hot flow structure (Perna, Raymond \& Narayan 2000)
are the subject of a separate study (Menou, Perna \& Raymond, in
preparation).

\section{Conclusion}

Motivated by the analogy between the optically-thin BL solutions of
Narayan \& Popham (1993) and Advection-Dominated Accretion Flows
(ADAFs), I constructed idealized models of ADAFs around WDs, for
application to quiescent DN. The critical assumptions made in this
work are discussed in detail in Appendix~A.

The Bremmstrahlung emission from the ADAF can account for the X-ray
emission properties of quiescent DN. In addition, if the energy
advected by the flow is thermalized in the WD atmosphere before being
radiated away, the strong HeII lines of many quiescent DN could
simply result from the photoionization of the disk
material by the UV-luminous WD.

The presence of an ADAF in quiescent DN could have several additional
interesting implications, such as offering a possible test of the
accretion flow geometry by the detection of an iron $K_{\alpha}$
fluorescence line from the X-ray irradiated disk, or providing a
possible explanation to the ``UV-delay'' of DN.

Although existing observational data already show a large diversity in
the DN population and indicate that ADAFs are probably not present in
all quiescent DN, prospects for testing the ADAF hypothesis in
promising DN candidates are good, especially with the recent advent of
the powerful X-ray satellites {\it Chandra} and {\it XMM-Newton}.

\section*{Acknowledgments}

I am grateful to Phil Charles, Eric Kuulkers and Tariq Shahbaz for
questionning the possible role of ADAFs around WDs, and to Ann Esin,
Brad Hansen, Jean-Pierre Lasota, Ramesh Narayan, Bohdan Paczynski,
Rosalba Perna, John Raymond and Eliot Quataert for useful discussions.
I thank R. Narayan for providing the code that he developed for
studying ADAFs around neutron stars. The calculations of the disk
reflection component make use of numerical routines developped by
A. Esin. Support for this work was provided by NASA through Chandra
Postdoctoral Fellowship grant number PF9-10006 awarded by the Chandra
X-ray Center, which is operated by the Smithsonian Astrophysical
Observatory for NASA under contract NAS8-39073.

\clearpage

\begin{appendix}
\section{Critical Assumptions}

The focus of this study is put on the emission spectrum of an ADAF
around a WD. In this appendix, I discuss a set of assumptions which
are crucial for the results presented. The validity of these
assumptions is not proven (nor obvious) and possible consequences of
their invalidity are discussed.  These assumptions concern the
structure of the hot flow surrounding the WD, the nature of the
interface between the flow and the WD, and the outcome of the energy
advected by the flow when it reaches the WD surface.

\subsection{The Hot Flow and the Interface with the WD}

Advection onto a WD differs significantly from accretion onto a BH in
that the gas does not have to cross a sonic point but, instead, it
must have a radial infall speed reaching zero at the stellar
surface. The neglect of this boundary condition, by using ADAF
dynamical solutions initially developed for the case of accretion onto
a BH (Popham \& Gammie 1998), is an important weakness of the present
study.  This description of the hot flow around the WD therefore
assumes a high radial infall speed for the gas (a fraction of its
free-fall speed) and a sudden reduction of this speed to zero only in
the close vicinity of the WD. Proving this assumption right or wrong
by solving the dynamical equations for the flow, with the proper
boundary conditions, is beyond the scope of the present work. It is
worth noting, however, that this sudden reduction of the radial infall
speed of the hot flow at the WD surface is found in the BL solutions
of Narayan \& Popham (1993).

Another region of the flow that is not described in detail in proposed
the models is the interface between the hot flow and the WD, where the
sudden reduction of the radial infall speed of the gas occurs. The
energetics of this region can be described in the following simple
terms.  In a thin accretion disk, half of the gravitational potential
energy available is radiated away by the disk, while the other half
remains in the flow in the form of rotational kinetic energy (Frank et
al. 1992). The energy budget is different in an ADAF, with three
energy reservoirs: the rotational kinetic energy, the radial (infall)
kinetic energy and the thermal kinetic energy of the hot gas (only a
small fraction of the available energy is radiated away by the flow).

In the models presented here,
\begin{eqnarray}
\Omega_{\rm ADAF}&\simeq&0.2 \Omega_K,\\
v&\simeq&\alpha_{\rm ADAF}V_{\rm ff} \simeq \alpha_{\rm ADAF} R \Omega_K,
\end{eqnarray}
where $\Omega_{\rm ADAF}$ is the rotation rate of the ADAF, $\Omega_K$
is the Keplerian rate, $v$ is the gas infall speed and $v_{\rm ff}$ is
the free-fall speed (see Narayan et al. 1998). This corresponds to rotational and kinetic energies per unit mass
\begin{eqnarray}
E_\phi&\simeq &\frac{1}{2} \left( 0.2 R \Omega_K \right)^2,\\ 
E_{\rm r}&\simeq &\frac{1}{2} \left( \alpha_{\rm ADAF} R \Omega_K \right)^2,
\end{eqnarray}
respectively. These energies are therefore comparable (for
$\alpha_{\rm ADAF}=0.2$), $\simeq 2 \times 10^{-2} R_{\rm WD}^2
\Omega_K^2(R_{\rm WD})$ per unit mass at the WD surface.

The total gravitational potential energy released per unit mass at the
WD surface is
\begin{equation}
E_{\rm pot} = \frac{GM_{\rm WD}}{R_{\rm WD}}=R_{\rm WD}^2
\Omega_K^2(R_{\rm WD}),
\end{equation}
so that $E_\phi$ and $E_{\rm r}$ are small, $\simeq 2 \times 10^{-2}
E_{\rm pot}$. The remaining part (i.e. most) of the potential energy
is in the form of thermal energy in the hot gas for an ADAF.

For the spectral models presented in this study, $E_\phi$ and $E_{\rm
r}$ have been added to the advected energy ($\sim E_{\rm pot}$) to
power the EUV radiation from the WD surface. The models shown in
Fig.~\ref{fig:two}a, for instance, have typical ratios of integrated
EUV luminosity to integrated X-ray luminosity of a hundred or more. If
the contributions $E_\phi$ and $E_{\rm r}$ were released in
optically-thin regions (at the interface between the ADAF and the WD
surface), they could modify the models predictions for the X-ray
emission significantly because they have magnitudes comparable to or
larger than the Bremsstrahlung emission from the hot flow. Note that
the non-zero rotation rate of the WD tends to reduce the importance of
$E_\phi$, especially for the values of a few \% to a few tens of \% of
$\Omega_K (R_{\rm WD})$ reported for the rotation rates of WDs in DN
(see, e.g., Sion 1999).

\subsection{The Fate of the Advected Energy}

Assuming that the properties of the ADAF surrounding the WD are those
discussed above, the fate of the energy advected by the flow when it
reaches the WD surface is still left unspecified. It is crucial for
the spectral predictions to know if this energy is released in
optically-thin or optically-thick regions at the WD surface.

There are two cooling channels for the hot gas carrying the energy:
conduction and radiation. I will now show that cooling by radiation
seems unlikely to result in the release of this energy in
optically-thick regions at the WD surface.

\subsubsection{Radiation}

Let's assume that the advected energy is indeed released in
optically-thick regions at the WD surface. The photospheric
temperature of the heated WD can then be estimated via
Eq.~(\ref{eq:one}), which yields
\begin{equation}
T_{\rm ph}=3.1 \times 10^4 \dot M_{16}^{1/4} M_{\rm WD}^{1/4}
R_9^{-3/4}~{\rm K},
\end{equation}
where $\dot M_{16}$ is the accretion rate onto the WD in units of
$10^{16}$~g~s$^{-1}$ and $R_9$ is the WD radius in units of
$10^9$~cm. We will assume $M_{\rm WD}=1.2 M_\odot$ and $R_9=0.5$ in
what follows. This gives
\begin{equation}
T_{\rm ph} \simeq 54,500~{\rm K},  
\end{equation}
for $\dot M_{16}=1$.

The density of the regions at the WD surface where the infalling of
gas in the ADAF is stopped can be estimated by equating the kinetic
pressure $P_{\rm gas}$ of the gas in these regions to the ram pressure
$P_{\rm ram} \equiv \rho v^2$ of the flow, with
\begin{eqnarray}
P_{\rm ram} &= &\frac{\dot M}{4 \pi R^2} \alpha_{\rm ADAF} V_{\rm ff}
	\\ & \simeq & 4.1 \times 10^5 \alpha_{\rm ADAF} \dot M_{16}
	M_{\rm WD}^{1/2} R_9^{-5/2}~{\rm dyne~cm^{-2}}
\end{eqnarray}
for the ADAF, where $V_{\rm ff} \equiv (2GM_{\rm WD}/R)^{1/2}$ is the
free fall velocity and $v=\alpha_{\rm ADAF} V_{\rm ff}$ is the radial
infall speed in the ADAF (Narayan \& Yi 1994).

The corresponding stopping density $\rho_{\rm stop}$ is therefore
given by
\begin{equation}
P_{\rm gas}=\frac{\rho_{\rm stop} k_{\rm B} T_{\rm p}}{\mu m_{\rm p}}
=P_{\rm ram},
\end{equation}
where I adopt a value $1/2$ for the mean molecular weight $\mu$,
$k_{\rm B}$ is the Boltzmann constant and $m_{\rm p}$ is the proton
mass. This gives
\begin{eqnarray}
\rho_{\rm stop} & \simeq & 7.9 \times 10^{-8} \alpha_{\rm ADAF} \dot
M_{16}^{3/4} M_{\rm WD}^{1/4} R_9^{-7/4}~{\rm g~cm^{-3}}\\ 
{\rm or} &
\simeq & 5.5 \times 10^{-8}~{\rm g~cm^{-3}}
\end{eqnarray}
for $\dot M_{16}=1$ ($M_{\rm WD}=1.2 M_\odot$ and $R_9=0.5$).  The
Rosseland mean opacity (see, e.g., Cox \& Tabor 1976) corresponding to
$\rho_{\rm stop}$ and $T_{\rm ph}$ is
\begin{equation}
\kappa_{\rm stop} \simeq 13~{\rm cm^2~g^{-1}}.
\end{equation}

Again, if the advected energy is released in optically-thick regions
of the WD atmosphere, the value of $\rho_{\rm stop}$ found above
should be at least comparable to the density at the photosphere of the
heated WD. For a standard Eddington atmosphere in hydrostatic
equilibrium, the photospheric density is given by the relation
\begin{equation}
P_{\rm ph} \simeq \frac{2 g_{\rm ph}}{3 \kappa},
\end{equation} 
where $P_{\rm ph}$, $g_{\rm ph}$ and $\kappa$ are the pressure,
gravity and opacity at the photosphere, respectively. The density at
the WD photosphere is therefore
\begin{equation}
\rho_{\rm ph} \simeq \frac{g_{\rm ph} \mu m_{\rm p}}{\kappa_{\rm stop}
k_{\rm B} T_{\rm ph}}.
\end{equation}
For $\mu=0.5$ and a WD with $M_{\rm WD} =1.2 M_\odot$ and $R_9=0.5$,
this yields
\begin{equation}
\rho_{\rm ph} \simeq 5.4 \times 10^{-6}~{\rm g~cm^{-3}},
\end{equation}
with a gravity $g_{\rm ph}=6.4 \times 10^8$~dyne~cm$^{-2}$.

The value of this photospheric density is two orders of magnitude
larger than the value $\rho_{\rm stop}$ derived above, which suggests
that the energy advected by the flow will not be released in
optically-thick regions at the WD surface if radiative processes act
alone.

\subsubsection{Conduction}

This still leaves the possibility that conduction processes
efficiently transport the advected energy deep in the WD atmosphere
where it is thermalized before being radiated away. The existing
literature on similar processes in the context of the post-shock
settling regions of WD accretion columns shows that it is difficult to
estimate the efficiency of these processes reliably. There are
proposed scenarios in which a large amount of the energy carried by
the hot gas behind the shock is buried deep into the WD atmosphere
(King \& Lasota 1979, 1980; Kuijpers \& Pringle 1982), but they have
been challenged (see, e.g., van Teeseling, Heise \& Paerels 1994;
Greeley et al. 1999).

The assumption of thermalization of the energy advected by the flow is
crucial to the spectral models for quiescent DN presented in this
study. If a fair fraction of the advected energy were not thermalized,
it would be radiated away as much harder emission from the vicinity of
the WD surface.  This would probably result in a soft (UV) component
(corresponding to a fraction of the hard emission that is reprocessed
by the WD atmosphere) comparable in luminosity to the hard (X-ray)
component itself. In that case, the strength of the HeII emission
lines could not easily be explained as photoionization of the disk
material by the central UV source, because it would probably be too
weak given the known level of X-ray emission of quiescent DN.

\end{appendix}

\clearpage

\clearpage

\clearpage

\begin{thebibliography}{}
\bibitem[]{}Abramowicz, M.A., Chen, X., Kato, S., Lasota, J.-P. \& Regev, O., 1995, ApJ, 438, L37.
\bibitem[]{}Belloni, T., et al., 1991, A\&A, 246, L44.
\bibitem[]{}Blandford, R.D. \& Begelman, M.C., 1998, MNRAS, 303, L1.
\bibitem[]{}Cannizzo, J.K. 1993, in Accretion Disks in Compact Stellar Systems, ed. J.C. Wheeler (Singapore: World Scientific), p. 6.
\bibitem[]{}C\'ordova, F.A. \& Mason, K.O., 1983, in Accretion Driven Stellar X-ray Sources, eds. W.H. Lewin \& E. van den Heuvel (CUP), p. 147. 
\bibitem[]{}Cox, A.N. \& Tabor, J.E., ApJS, 31, 271.
\bibitem[]{}Eracleous, M., Halpern, J. \& Patterson, J., ApJ, 382, 290 (1991).
\bibitem[Esin et al. (1997)]{esietal97}Esin, A.A., McClintock, J.E., \& Narayan, R., 1997, ApJ, 489, 865.
\bibitem[]{}Esin, A.A., Narayan, R., Cui, W., Grove, J.E. \& Zhang, S.-N., 1998, ApJ, 505, 854.
\bibitem[]{}Frank, J., King, A. \& Raine, D., 1992, Accretion Power in Astrophysics (Cambridge: Cambridge University Press).
\bibitem[]{}Greeley, B.W., Blair, W.P., Long, K.S. \& Raymond, J.C., ApJ, 513, 491.
\bibitem[]{}George, I.M. \& Fabian, A.C., 1991, MNRAS, 249, 352.
\bibitem[]{}Hameury, J.-M., Lasota, J.-P. \& Dubus, G., 1999, MNRAS, 303, 39.
\bibitem[]{}Hameury, J.-M., Lasota, J.-P., McClintock, J. E. \& Narayan, R., 1997, ApJ, 489, 234.
\bibitem[]{}Ichimaru, S., 1977, ApJ, 214, 840.
\bibitem[]{}King, A.R. \& Lasota J.-P., 1979, MNRAS, 188, 653.
\bibitem[]{}King, A.R. \& Lasota J.-P., 1980, MNRAS, 191, 721.
\bibitem[]{}Kippenhahn, R. \& Thomas, H.-C., 1978, A\&A, 63, 265. 
\bibitem[]{}Kuijpers, J. \& Pringle, J.E., 1982, A\&A, 114, L4.
\bibitem[]{}Lasota, J.-P., Narayan, R. \& Yi, I., 1996, A\&A, 314, 813.
\bibitem[]{}Liu, B.F., Meyer, F. \& Meyer-Hofmeister, E., 1997, A\&A, 328, 247.
\bibitem[]{}Lightman, A.P. \& White, T.R., 1988, ApJ, 335, 57.
\bibitem[]{}Long, K., 1996, in IAU Colloquium 158, ``Cataclysmic Variables and Related Objects'', Eds. A. Evans \& J.H. Wood, p. 233.
\bibitem[]{}Mauche, C.W., 1996, in Astrophysics in Extreme Ultraviolet, IAU Coll. 152, Bowyer, S. \& Bowyer, R.F., eds, (Kluwer: Dordrecht), p. 317.
\bibitem[]{}Menou, K. \& McClintock, J.E., 2000, ApJ, submitted.
\bibitem[]{}Meyer, F. \& Meyer-Hofmeister, E., 1994, A\&A, 288, 175.
\bibitem[]{}Morrison, R. \& McCammon, D., 1983, ApJ, 270, 119.
\bibitem[]{}Mukai, K. \& Shiokawa, K., 1993, ApJ, 418, 863.       
\bibitem[]{}Mukai, K., Wood, Janet H., Naylor, T., Schlegel, E.M. \& Swank, J.H., 1997, ApJ, 475, 812.
\bibitem[]{}Narayan, R., Barret, D. \& McClintock, J.E., 1997, ApJ, 482, 448.
\bibitem[]{}Narayan, R., Mahadevan, R. \& Quataert, E., 1998b, in The Theory of Black Hole Accretion Discs, eds. M. A. Abramowicz, G. Bjornsson, and J. E. Pringle (Cambridge: Cambridge University Press), Astro-ph/9803141.
\bibitem[Narayan et al. (1995)]{mny95}Narayan, R., McClintock, J.E. \&
Yi, I., 1996, ApJ, 457, 821.
\bibitem[]{}Narayan, R. \& Popham, R., 1993, Nature, 362, 820.
\bibitem[]{}Narayan, R. \& Raymond, J.C., 1999, ApJL, 515, L69.  
\bibitem[]{}Narayan, R. \& Yi, I., 1994, ApJ Lett., 428, L13.
\bibitem[]{}Narayan, R. \& Yi, I., 1995, ApJ, 444, 231.
\bibitem[]{}Orosz, J.A., Remillard, R.A., Bailyn, C.D. \& McClintock, J.E., 1997, ApJ, 478, L83. 
\bibitem[]{}Patterson, J. \& Raymond, J.C., 1985a, ApJ, 292, 535.
\bibitem[]{}Patterson, J. \& Raymond, J.C., 1985b, ApJ, 292, 550.
\bibitem[]{}Perna, R., Raymond, J.C. \& Narayan, R., 2000, ApJ, in press, Astro-ph/0005387.
\bibitem[]{}Popham, R. \& Gammie, C.F., 1998, ApJ, 504, 419.
\bibitem[]{}Pratt, G.W., Hassall, B.J.M., Naylor, T. \& Wood, J.H., 1999, MNRAS, 307, 413.
\bibitem[]{}Pringle, J.E. \& Savonije, G.J., 1979, MNRAS, 187, 777.
\bibitem[]{}Quataert, E. \& Gruzinov, A., 1999, ApJ, 520, 248.
\bibitem[]{}Quataert, E. \& Narayan, R., 1999a, ApJ, 516, 399.
\bibitem[]{}Quataert, E. \& Narayan, R., 1999b, ApJ, 520, 298.
\bibitem[]{}Rees, M.J., Phinney, E.S., Begelman, M.C. \& Blandford, R.D., 1982, Nature, 295, 17.
\bibitem[]{}Sion, E.M., 1999, PASP, 111, 532.
\bibitem[]{}Szkody, P., Long, K.S., Sion, E.M. \& Raymond, J.C., 1996, ApJ, 469, 834.
\bibitem[]{}Tylenda, R., 1981, Acta Astr., 31, 267. 
\bibitem[]{}van Teeseling, A., Beuermann, K. \& Verbunt, F., 1996, A\&A, 315, 467.
\bibitem[]{}van Teeseling, A., Heise, J. \& Paerels, F., 1994, A\&A, 281, 119.
\bibitem[]{}Vrtilek, S.D., Silber, A., Raymond, J.C. \& Patterson, J., 1994, ApJ, 425, 787.
\bibitem[]{}Warner, B., 1995, Cataclysmic Variable Stars, (Cambridge: Cambridge University Press).
\bibitem[]{}Wheeler, J.C., 1996, in Relativistic Astrophysics, eds. B. Jones \& D. Markovic (Cambridge: Cambridge University Press), p. 211.
\bibitem[]{}Wood, J.H., Naylor, T., Hassall, B.J.M. \& Ramseyer, T.F., 1995, MNRAS, 273, 772.
\bibitem[]{}Yoshida, K., Inoue, H. \& Osaki, Y., 1992, PASJ, 44, 537.

\end{thebibliography}
\end{document}